\documentclass[11pt,twoside]{article}

\usepackage{asp2004}
\usepackage{epsf}
\usepackage{psfig}
\usepackage{lscape}
\usepackage{graphicx}

\begin{document}
\title{Massive Stars in the SMC}   
\author{D.J. Lennon}   
\affil{Isaac Newton Group of Telescopes, Apartado 321, E-38700 Santa Cruz 
de La Palma, Spain; Instituto de Astrofisica de Canaries, E-38205 La Laguna,
Tenerife, Spain}    
\author{C.J. Evans}
\affil{Isaac Newton Group of Telescopes, Apartado 321, E-38700 Santa Cruz 
de La Palma, Spain}
\author{C. Trundle}
\affil{Instituto de Astrofisica de Canaries, E-38205 La Laguna,
Tenerife, Spain}

\begin{abstract} 
In this paper we discuss how the Small Magellanic Cloud is the ideal
laboratory in which to study massive stars in a low metallcity 
environment. We review the observational data for OB stars in the 
SMC concentrating on those aspects of their spectra
which provide information on processes which may
strongly influence their evolution, namely mass-loss,
rotational mixing and mass-transfer.  We illustrate the
very weak winds now thought to pertain to late O-dwarfs 
in the SMC, using HST/STIS observations of the main sequence
in the very young cluster NGC\,346, briefly discussing the
quantitative results for these stars, and the difficulties
involved in their determination. We show how stars with similar 
luminosities can have different luminosity classes while
stars with similar spectral types and luminosity classes can have significantly
different luminosities. These discrepancies can be interprated as evidence for 
rotational mixing on the main sequence. While the weak winds of the 
dwarfs present serious difficulties for the determination of wind terminal 
velocities we show that the supergiants have terminal velocities 
comparable to OB supergiants in the Milky Way, in reasonable agreement
with theory.  We also summarize recent work demonstrating that the
temperature dependence of wind terminal velocities does not follow
the widely adopted step-like approximation, the bistability
jump around spectral type B1 does not occur for normal stars.
Finally we review surface compositons of OB stars in the SMC finding
that 42 out of 45 OB stars with detailed surface abundances are enriched
in nitrogen by a factor $\sim 10$ or more. While these enhancements
are consistent with those produced by models with rotational mixing 
the rotational velocities of the sample are significantly lower than
the values predicted by the modesls,  indicating a possible problem with
the evolution of angular momentum in the models or possibly in the 
efficiency of mixing.  In this context we comment on the biases present in
the stellar samples discussed in the literature.
\end{abstract}



\section{Introduction}

\begin{figure}[!ht]
\plotone{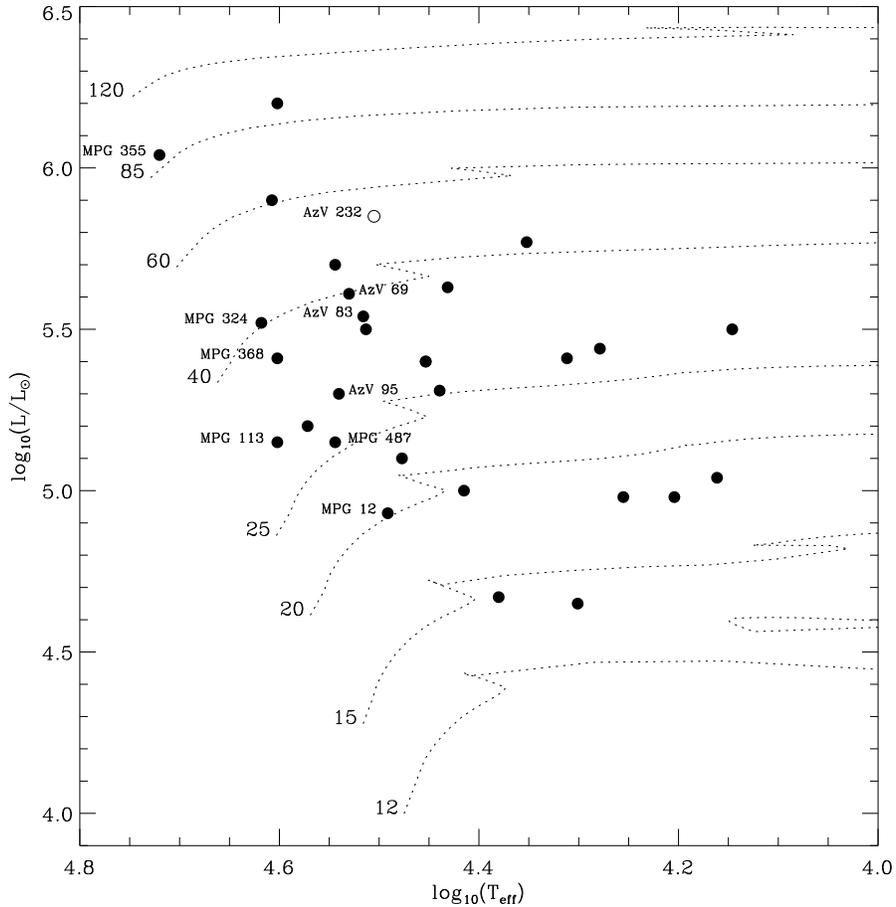}
\caption{HR diagram illustrating the position of the SMC stars observed in
HST/STIS GO programmes 7437 and 9116 (filled cirlces).  The evolutionary tracks are from
Charbonnel et al (1993) for approximate SMC metallicity and without 
rotation.  Luminosities and effective temperatures are taken
from the sources discussed in Evans et al (2004b).  The stars labelled
with MPG numbers (Massey et al 1989) are all in the SMC cluster NGC\,346,
their spectra are illustrated in Fig.2.  Those labelled with their
AzV numbers are the O7 supergiants and giants discussed in section 3 along
with the additional O7\,Iaf+ star AzV\,232 (open symbol).
}
\end{figure}

The Small Magellanic Cloud (SMC) has a special significance in
the study of massive stars due to the fact that its metallicity (Z) is
approximately $1/5^{\rm th}$ solar.  This importance stems from the fact
that mass-loss rates scale very roughly as $\sqrt{Z}$ which implies that
we should see significant differences in mass-loss rates of SMC OB 
stars with respect to similar stars in the solar neighbourhood.
The crucial corollary to this is the fact that mass-loss is a major 
factor influencing massive star evolution, and if we also consider 
suggestions that stellar rotation may well depend on Z, it is clear that
the SMC is the ideal laboratory in which to study massive stars 
and their evolution
at low Z.  

There are additional practical reasons which make the SMC
the ideal environment in which to carry out these studies: The well determined
distance of the SMC enable us to deduce accurate stellar radii, which are
important in the application of unified model atmosphere/wind models, and
allow us to deduce spectroscopic masses for comparison with predictions
of stellar evolution calculations.  The low extinction towards the SMC gives us access
to the wind diagnostics in the near- and far-ultraviolet regions of the
spectrum, and paticularly allows us to determine wind terminal velocities
from saturated P-Cygni lines.  It is perhaps not generally appreciated that
due to extinction it is impractical to observe all but a handful of OB stars
in our galaxy in the FUV.   Finally, the nitrogen abundance of
the SMC is much lower than the mean metal deficiency in the SMC, being
only about $1/30^{\rm th}$ solar.  This permits an easy detection
of surface nitrogen anomalies from different processes such as mass-loss,
mixing and mass transfer.
The objective of this review is therefore to look at recent observational 
programmes which concentrated on the SMC and highlight some results concerning
mass-loss, stellar rotation and surface compositions of massive stars in that
galaxy.

Perhaps the definitive ultra-violet spectroscopic survey of massive stars in the
SMC to date is represented by the survey discussed
by Walborn et al  (2000) and Evans et al (2004b), the sample's coverage
of the HR diagram being illustrated in Fig.1.  These data were
obtained in two separate Hubble Space Telescope (HST) programmes, GO7437 concerning
O-type stars, and GO9116 concerning B-type stars, the medium
resolution echelle mode of the Space Telescope Imaging Spectrograph 
(STIS) was used for both programmes. A general
consideration for the target selection was to distribute targets
throughout the upper HR diagram but ensuring that we obtained
spectra of stars close to the previously unexplored low
luminosity O-type dwarfs.  This latter constraint was achieved by
targetting O-type stars in the very young SMC cluster NGC346, while low 
luminosity giants were observed in the rather older cluster NGC330,
other giants and supergiants being chosen from the catalogues of 
Azzopardi \& Vigneau (1975, 1982),  though using the spectral types of 
Garmany et al (1987), Massey et al (1989), Lennon et al (1993) 
and Lennon (1997).  In choosing the targets it was noticed that there was an
O7Iaf+ star (AzV 83) with a visual magnitude in the range occupied
by O7\,III stars and this was added to the target list together with an O7\,III
star (AzV 69) of similar magnitude.  In our subequent discussion of
the observed properties of SMC massive stars we concentrate on
results derived from this sample, although we include some results from other
notable observational programmes.  We only briefly discuss mass-loss, since a
more thorough discusson of this topic is provided elsewhere in these proceedings
via the contributions of Alex de Koter and Trundle et al.

\section{O-stars with weak winds}

The exceptional weakness of stellar wind features in the SMC O-type dwarfs 
was first illustrated and discussed by Walborn et al (1995), this being
enhanced by the Walborn et al (2000) study, referred to above, which
extended coverage of the HR diagram to lower masses and luminosities than
had previously been achieved.  The O-type dwarf sequence, as represented
by the stars in NGC346, is a particularly good illustration of the weak
wind features in SMC stars (Fig.2), the late O-type stars exhibiting almost
no sign of the characteristic P-Cygni wind features usually associated with
O-type stars.  Optical spectra of these objects confirm the
weakness of their winds in the the lack of significant in-filling or emission of the 
H$\alpha$ line, usually a primary diagnostic of the mass-loss rate. 

\begin{figure}[!ht]
\centerline{\psfig{figure=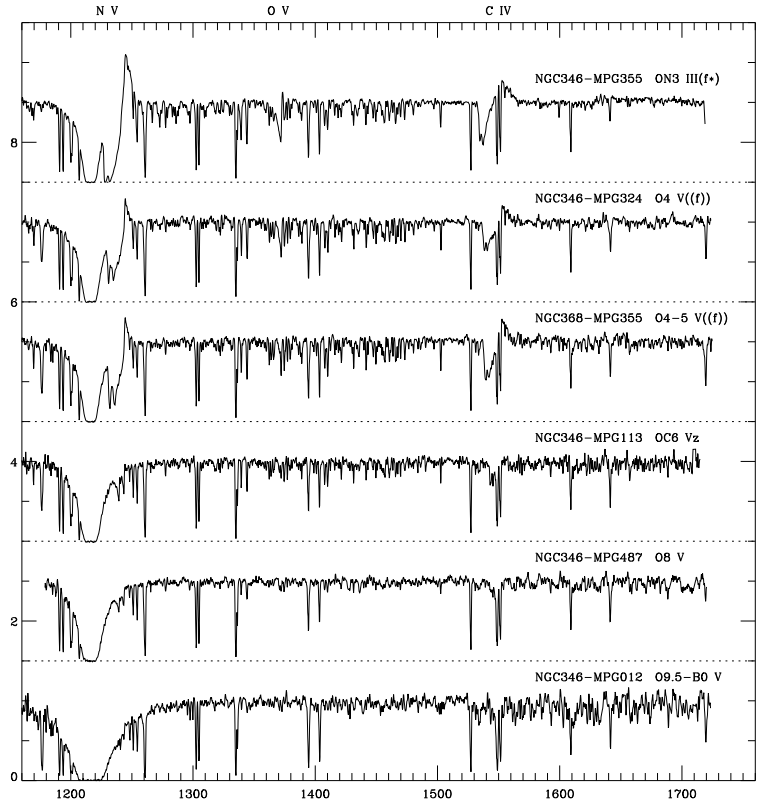,height=11cm}}
\caption{Illustration of the HST/STIS UV spectra for the 
O-type dwarf sequence in the SMC cluster NGC346. Notice how the
P-Cygni wind features of the N\,{\sc v} and C\,{\sc iv} lines 
fade away towards the later spectral types.}
\end{figure}

This theme was followed up by Bouret et al (2003) who determined mass-loss
rates for the NGC346 stars using CMFGEN to model the $weak$ wind features
in their UV spectra.  The surprising result was that the lower luminosity
dwarfs MPG113 (OC6\,Vz), MPG487 (O8\,V) and MPG12 (O9.5-B0\,V (N str)) have
mass-loss rates which are one to two orders of magnitude lower than those
predicted by the theory of Vink et al (2001), Fig.3. Independent corroboration
for this result was provided by Martins et al (2004) who analysed the spectra
of a further 4 SMC dwarfs and found similar results (also illustrated in Fig.3).
It is not clear why low metallicity O-type dwarfs appear to have such
weak winds, obviously clumping only exacerbates the problem.
A major problem in the interpretation
is that the winds are so weak that the H$\alpha$ line in these stars, 
normally the most important mass-loss diagnostic, is almost a pure photospheric
profile and contains essentially no information on the mass-loss rate.
The UV P-Cygni profiles are also very weak, and as demonstrated by  the
Martins et al work, often only upper limits on the mass-loss rate are obtained.
Clearly a more sensitive mass-loss diagnostic is required, one possibility
being the use of the Br$\alpha$ line. 

\begin{figure}
\centerline{\psfig{figure=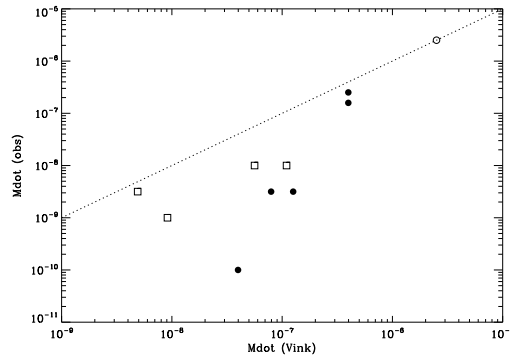,height=5cm}}
\caption{Comparison of observed and theoretical mass-loss rates for O-type
dwarfs (filled cirles) and giants (open circles) in  the SMC cluster NGC346,
from Bouret et al (2003). The open squares are the upper-limits derived
by Martins et al (2004) for dwarfs in SMC-N81. 
The dotted line represents a one-to-one
correspondence between observation and theory, note how the
dwarfs appear to have mass-loss rates which are one to two orders
of magnitude lower than the theoretical predictions of Vink et al (2001).}
\end{figure}

\section{O7 giants and supergiants}

As pointed out by Walborn et al (2000), AzV\,83 is one of only two normal
Of stars in the SMC (both are of O7\,Iaf+ spectral type), 
the other being the well known star AzV\,232 = Sk\,80
(which is associated with NGC346). However while 
their optical and visual spectra are virtually identical 
AzV\,83 is 0.9 mag fainter in $M_v$.  In fact, as already alluded to
above, the magnitude of AzV\,83 is lies in the range normally occupied
by O7 giants.  Indeed Walborn et al pointed out that it occupied an
almost identical position in the HR diagram to the OC7.5\,III giant AzV\,69
and speculated that the latter was slow rotator while AzV\,83 was initally
a fast rotator on the main sequence which has evolved to higher luminosities
and now coincides with the position of the more massive giant AzV\,69.
Support for this scenario is provided by the OC classification of AzV\,69
which might indicate that its carbon abundance is primeval while that
of AzV\,83 enhanced.  Walborn et al discussed the additional comparison of
AzV\,69 with the normal O7\,III giant AzV\,95 which provide a striking
contrast of N/C line strengths and infer that nitrogen is enhanced in
AzV\,95 relative to AzV\,69, again supporting the scenario in which
the latter is unaffected by rotation.

\begin{table}[!ht]
\caption{Comparison of properties of AzV\,69 and AzV\,83 from 
Hillier et al (2003) and Walborn et al (2000). Note the different
clumping factors ($f$), spectroscopic masses and surface nitrogen
abundances.}
\smallskip
\begin{center}
{\small
\begin{tabular}{ccccccccc}
\tableline
\noalign{\smallskip}
Star & Sp. type & $V$  & $v\sin i$ & log\,g & Mass  &
\.{M} & $f$ &  [N/N$_{\odot}$]  \\
 & & mag. & km/s & & ($M_{\odot}$) & ($M_{\odot}$/yr) &  &  \\
\noalign{\smallskip}
\tableline
\noalign{\smallskip}
AzV\,69 & OC7.5\,III((f)) & 13.35 & 70 & 3.50 & 40 & 9.2x10$^{-7}$ & 1.0 &
0.02\\
AzV\,83 & O7\,Iaf+ & 13.58 & 80 & 3.25 & 22 & 7.3x10$^{-7}$ & 0.1 & 1.8 \\
\noalign{\smallskip}
\tableline
\noalign{\smallskip}
\end{tabular}
}
\end{center}
\end{table}

These qualitative comparisons between AzV\,69 and AzV\,83
were investigated in detail by Hillier et al (2003)
who used CMFGEN to analyse their UV and optical spectra.
They found that both stars have similar values of effective temperatures
(approximately 32\,000\,K), $v\sin i$, luminosity, and mass-loss rate
although the UV spectrum of AzV\,83 was best reproduced assuming a
clumping factor of $f=0.1$ compared to 1.0 for AzV\,69.  However it
was also found that AzV\,83 relative to AzV\,69 has a lower surface
gravity and consequently a lower spectroscopic mass, while the former
has a surface nitrogen abundance of approximately twice solar compared
with a near normal SMC nitrogen abundance for the latter.  This lends
strong credence to the idea expresed above that AzV\,83 was initially a fast
rotator but has undergone rotationally enhanced mass-loss and mixing resulting
in a lower current mass and higher surface nitrogen abundance than AzV\,69.
Moreover, AzV\,69 must have a low rotational velocity (not just $v\sin i$)
since it has evolved away from the zero-age main sequence (ZAMS) but has
pristine surface composition for the SMC.

\section{Wind terminal velocities}

Radiation driven wind theory predicts rather simple dependence of the
ratio of wind terminal velocity to escape velocity such that 
$v_{\infty}/v_{esc} \sim \hat{\alpha}/1-\hat{\alpha} $ where $\hat{\alpha}$
is the effective value of the force multiplier parameter $\alpha$ 
defined by Puls et al (2000). Puls et al show that the value of
$\hat{\alpha}$ is rather insensitive to metallicity, at least in the 
range 0.1 to 3.0 times solar.  Hence one expects that to first order
the wind terminal velocities in the SMC should be rather similar to
those of galactic stars.  This is somewhat at odds with the oft-stated
comment that wind terminal velocities in the SMC are lower than those
in the Galaxy.  Evans et al (2004a) have carefully investigated this
issue using a well controlled sample of stars in the SMC which
have saturated P-Cygni UV line profiles and well determined escape
velocities.  They find (Fig.4) that there is no significant difference between
SMC and Galactic samples but it should be noted that the requirement
for stars to have saturated UV resonance lines implies that the sample
consists of supergiants or giants, but not dwarfs (due to the weaker
winds in the SMC).  Nevetheless, this result is consistent with the predictions
of Puls et al and lends support to the assumption that
wind terminal velocities supergiants in Local Group galaxies are
similar to those of galactic stars (Urbaneja et al 2003).

\begin{figure}[!ht]
\includegraphics[scale=0.5]{lennon_fig4a.eps}
\includegraphics[scale=0.38]{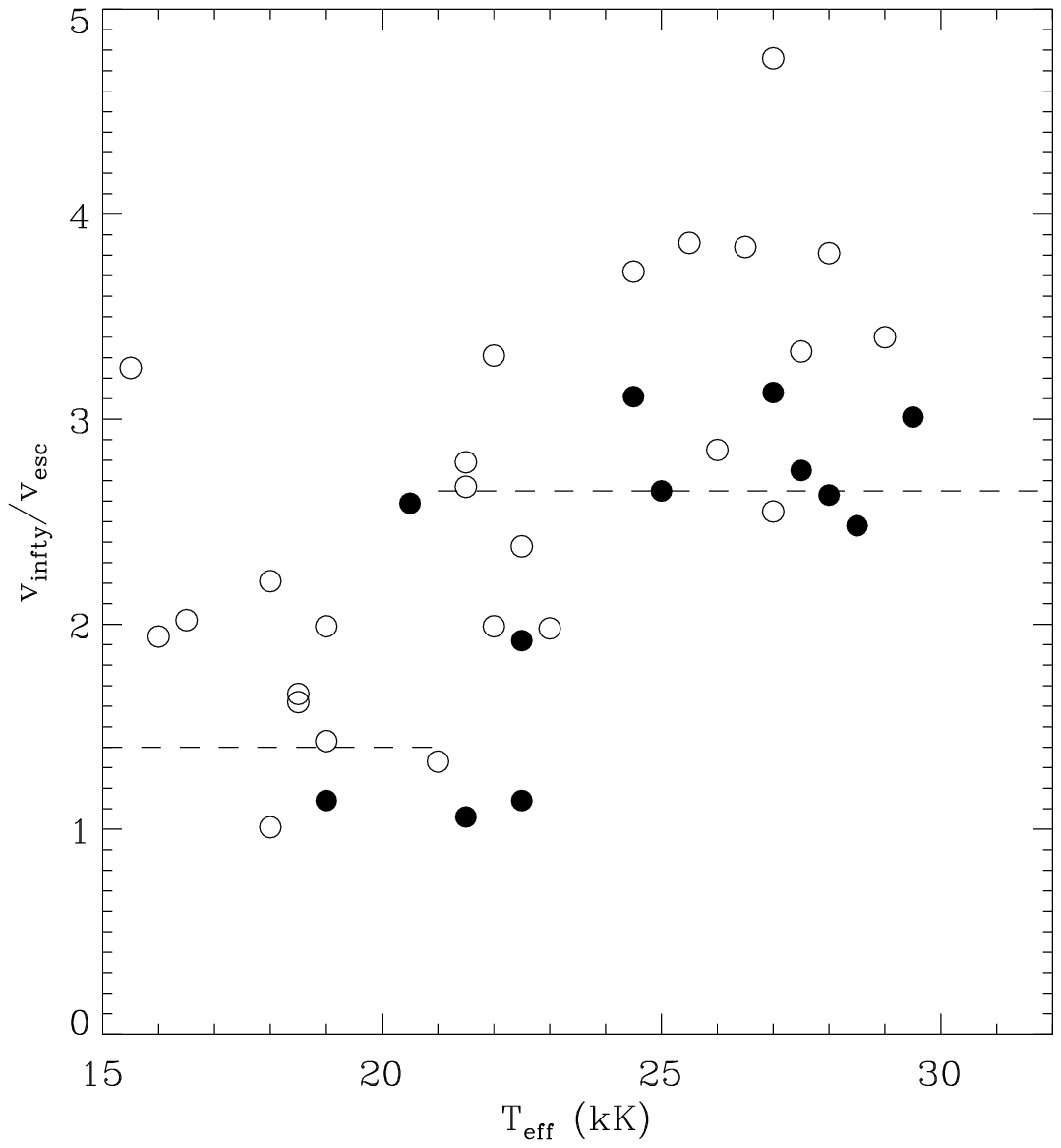}
\caption{The left-hand panel illustrates the results of Evans et al (2004a) 
comparing the ratio $v_{\infty}/v_{esc}$  for OB stars in the SMC (filled
circels) with Galactic results (open circles).
The right-hand panel is from Crowther et al (2005) and shows
improved Galactic results in the vicinity of the bistability jump at
21\,000\,K. In both panels the horizonal lines represent the scalings
of Kudritzki \& Puls (2000).  These figures show the large scatter in
Galactic values due to uncertain distances, similar values for SMC and
Galactic stars, and poor agreement with the assumed scalings of Kudritzki
\& Puls.}
\end{figure}

We now turn to the effective temperature dependence of terminal
velocities.  Lamers et al (1995) and Kudritzki \& Puls (2000) have shown
that the ratio $v_{\infty}/v_{esc}$ depends on effective temperature
and argue that a step function, with break at early B-type stars, 
is a good approximation to the observed trend.  This step has been interprated
as a manifestation of the bistability effect in LBV winds,
while the calibration of Lamers et al is widely used, for example it is
adopted in the theoretical estimation of mass-loss rates (Vink et al 2001) 
and in the formation of circumstellar neulae and, for example, their impact on GRB
light curves. We can see from Fig.4 (left panel) that the latest 
observational data do not lend strong support to this calibration, the 
galactic data in particular have significant scatter due mainly to the uncertain
distance to galactic OB stars.  Crowther et al (2005) have looked in more
detail at the B-type bistability jump at 21\,000\,K (Fig.4, right-hand panel)
and show that for galactic B-type supergiants the trend is more gradual than
step-like.  Note also that these new results for Galactic supergiants are in
good agreement with the SMC results of Evans et al (2004a). 

\section{Rotation and surface composition} 

The surface composition of a massive star may be modified during its evolution
by mass-loss, rotationally induced mixing, mass-transfer (in binary systems) and
magnetic fields.  We do not consider the very important topic of massive binary 
evolution here, but consider only predictions of those calculations for single star
evolution which include mass-loss and rotation, the most obvious spectroscopic consequence 
of which is an enhanced surface nitrogen abundance.    Since the pristine nitrogen 
abundance of the SMC is  $1/30^{\rm th}$ solar 
evolutionary effects on the surface nitrogen abundance of massive stars are
much easier to detect in the SMC than in our Galaxy or the LMC. 
Fig.5 represents a compilation of the surface nitrogen abundances for O and
B-type stars in the SMC.  This sample contains results for all the objects
shown in Fig.1, (Lennon et al 2003, Bouret et al 2003, Hillier et al 2003,
Trundle et al 2004, Heap et al 2005, Evans et al 2004c)
with additional results from recent work on B-type supergiants
in the SMC (Trundle \& Lennon 2005, Dufton et al 2005). The evolutionary tracks
are from Maeder \& Meynet (2001) for an initial rotational velocity of 300 km/s.

\begin{figure}[!ht]
\centerline{\psfig{figure=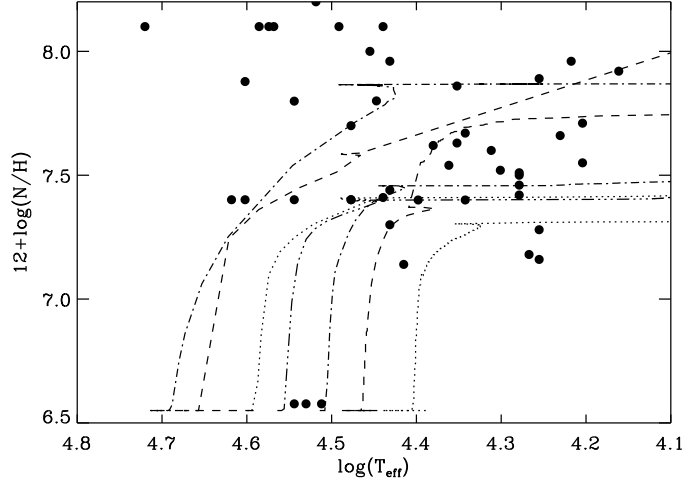,height=7cm}}
\caption{Surface nitrogen abundances of SMC OB stars as a function
of effective temperature compared to the predictions of the Maeder \& 
Meynet (2001) calculations for an initial rotational velocity of 300 km/s and for
initial masses 60, 40, 25,20, 15, 12 and 9 solar masses. Typical error
estimates for stars with log$(T_{eff}) < 4.5$ are $\pm 0.2$.  The pristine
SMC nitrogen abundance is taken to be 6.55 dex. }
\end{figure}

One can see that most OB stars show enrichments in nitrogen by a factor of
10 or more, comparable to that exhibited by the evolutionary tracks at the end
of the main sequence.  For the O9/B-type stars, those with log$(T_{eff}) < 4.5$, 
this is what one might expect since these objects are all evolved stars (refer to Fig.1).
For O-type stars, while it appears that the nitrogen enrichments are rather higher 
than one  might expect from the evolutionary tracks, one must bear in mind that
there are few quantitative error bars for the O-type stars (these are typically 0.2 dex 
for the O9/B-type objects).  These abundances are typically derived from spectrum
synthesis fits, clearly more quantitative estimates are desired.  
However there is a more serious problem concerning Fig.5 which is that the
mean $v\sin i$ for the B-star and O-star samples are 65 and 67 km/s respectively,
much lower than the initial rotational velocity of the models, and the
terminal age main sequence rotational velocities of $\sim 200$ km/s.  This problem
has been discussed by Herrero \& Lennon (2004) in the context of Galactic 
OB stars, and may point to a problem with the angular momentum evolution
of the models.  One must bear in mind however that since we cannot derive 
reliable metal abundances for stars with $v\sin i$ greater than $\sim 120$ km/s
we are biased towards stars with low $v\sin i$, and hence also low $v$. 

In an interesting investigation of two pole-on Be stars in the SMC cluster
NGC330, Lennon et al (2005) showed that there is strong evidence that
they are not enriched in nitrogen raising the interesting possibility
that rotational mixing is not an efficient process for stars in this
mass range ($\sim 10\,M_{\odot}$). While one might speculate about the inhibiting
influence magnetic fields in massive stars, stellar evolution calculations
are so far exploratory (Maeder \& Meynet 2004) and quantitative predictions of 
surface abundances for such models are lacking.

Finally we note that Maeder et al (1999) raised the interesting 
possibility that stellar rotation depends on metallicity, based
on the observed dependence of the fraction of Be stars in clusters
in the Galaxy, the LMC and the SMC.   This idea was strongly driven
by the high fraction of Be stars found in the SMC cluster NGC330 of 40\%,
together with a perceived low fraction of Be stars in clusters in the
solar neighbourhood.  However Pigulski \& Kopacki (2000) have found a
similarly high number of Be stars of $36\pm 7$\% in the cluster NGC7419 
which is 2 kpcs towards the galactic centre indicating that there exists
large statistical fluctuations in the Be star fraction at a given Z.
Penny et al (2004) concluded that there was no dependence on Z for O-type
stars in the Galaxy, LMC and SMC but their samples for the LMC and SMC
consisted of only 19 and 15 stars respectively and suffer from the same
bias to narrow-line stars discussed in the previous paragraph. 
Furthermore  Keller (2004) investigate $\sim 100$
B-type cluster and field stars in the LMC and concluding that there is mild
evidence for faster rotational velocities than their galactic counterparts.
Clearly, the task of correlating abundance anomalies with rotational
velocities requires the use of larger unbiased samples, such a project is discussed
by Evans et al elsewhere in these proceedings.

\acknowledgements 
DJL would like to thank all the participants of GO7437 and 9116 including Rolf Kudritzki,
Nolan Walborn, Joachim Puls, Adi Pauldrach, Linda Smith, Sally Heap, Max Pettini, Steve
Smartt, Thiery Lanz, Ivan Hubeny, Norbert Langer, Claus Leitherer, Joel Parker, Timothy Heckman.
We also acknowledge support of PPARC under grant PPA/G/S/2001/00131. 



\end{document}